# Electronic origin of the incommensurate modulation in the structure of phosphorus IV

**V F Degtyareva**

Institute of Solid State Physics Russian Academy of Sciences, Chernogolovka, Russia

degtyar@issp.ac.ru

**Abstract**. An incommensurate modulated structure was found recently in a light group V element phosphorous in the phase P-IV stable in the pressure range $107 - 137$ GPa. We consider configurations of the Brillouin zone and Fermi sphere within a nearly-free-electron model in order to analyze the importance of these configurations for the crystal structure energy. For the phase P-IV with the base-centered orthorhombic structure, $oC2$, we consider a commensurate approximant with an 11- fold supercell along the c-axis and a modulation wave vector equal 3/11 which is close to the experimentally observed value of 0.267. Atomic shifts due to the modulation result in appearance of satellite reflections and hence in a formation of additional Brillouin zone planes. The stability of this structure is attributed to the lowering of the electronic band structure energy due to Brillouin zone – Fermi surface interactions.

## 1. Introduction

Recent high-pressure x-ray diffraction studies revealed unusual complex and low symmetry structures in some simple elements [1], including incommensurate modulated (IM) structures. IM structures were found in the elements of group VII (I and Br) and group VI (S, Se and Te). Very recently, an IM structure was found in a light group V element phosphorous in the phase P-IV stable in the pressure range $107 - 137$ GPa [2]. This phase is intermediate between simple cubic and simple hexagonal structures that have atomic coordination equal 6 and 8, respectively. All IM structures were observed when elements become metallic under pressure. This implies the importance of the two main contributions to the lattice energy: electrostatic (Ewald) and electronic (band structure) energies. The latter can be lowered due to a formation of a Brillouin zone plane and an opening of an energy gap at this plane. This effect is important for stability of so-called Hume-Rothery phases in Cu – Zn and related alloys [3]. Under pressure, the band structure energy part becomes more important leading to a formation of complex low-symmetry structures [4].

For a classical Hume-Rothery phase $Cu_5Zn_8$, the Brillouin zone filling by electron states is equal to 93%, and is around this number for many other phases stabilized by the Hume-Rothery mechanism [5]. Diffraction patterns of these phases have a group of strong reflections with their scattering vectors $q_{hkl}$ lying near $2k_F$ and the Brillouin zone planes corresponding to these $q_{hkl}$ form a polyhedron that is very close to the Fermi sphere. We consider configurations of the Brillouin zone and Fermi sphere within a nearly-free-electron model in order to analyze the importance of these configurations for the lattice energy. A computer program, BRIZ, has been developed for the visualization of Brillouin zone – Fermi sphere configurations to estimate Brillouin zone filling by electron states [5]. Thus, with the BRIZ program one can obtain a qualitative picture and some quantitative characteristics on how a structure matches the criteria of the Hume-Rothery mechanism.

## 2. Results and Discussion

The phase P-IV has an incommensurately modulated crystal structure with superspace group *Cmmm* (00γ) *s*00 [2,6], with the modulation wave vector γ = 0.2673. The basic lattice is base-centered orthorhombic with lattice parameters $a = 2.772$ Å, $b = 3.215$ Å, $c = 2.063$ Å at 125 GPa [2]. Diffraction pattern and suggested structural model are shown in Figure 1. We consider a commensurate approximant with an 11-fold supercell along the c-axis and a modulation wave vector equal to $3/11 = 0.273$ which is close to the experimentally observed value of 0.267.

Taking the experimentally found atomic volume 9.19 Å$^3$ and the number of valence electrons for group V element P equal $z = 5$ we estimate the value of $2k_F$ equal to 5.05 Å$^{-1}$. Within a nearly free-electron model the Fermi sphere radius is defined as $k_F = (3\pi^2 z/V)^{1/3}$, where $z$ is the number of valence electrons per atom and $V$ is the atomic volume. The $2k_F$ position is indicated on Figure 1 (right inset). Atomic shifts due to the modulation result in appearance of satellite reflections and hence in a formation of additional Brillouin zone planes. Reflections have scattering vectors H = $h$a*+ $k$b*+ $l$c*+ $m\gamma$c* = $h$a*+ $k$b*+ ($l$+$m\gamma$)c*. For the commensurate approximant, $\gamma = 3/11$, an 11-fold supercell has been employed.

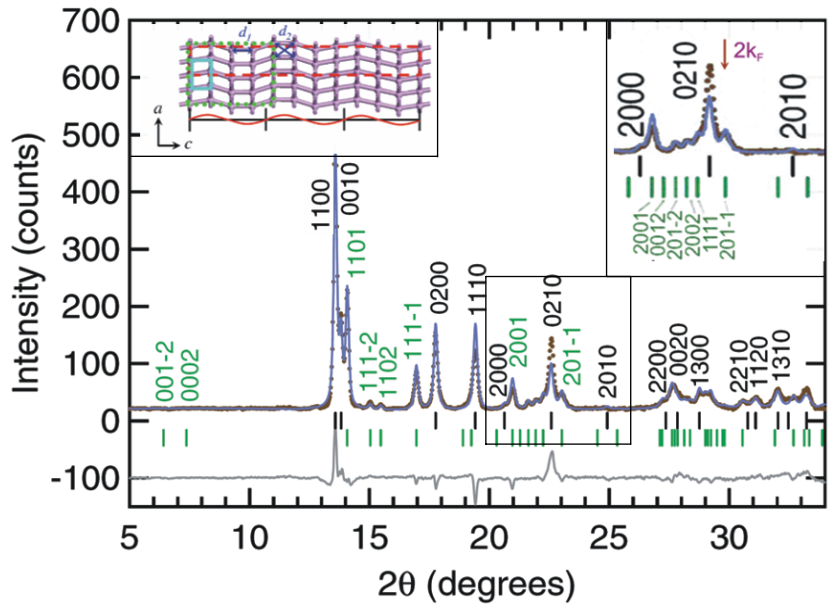

**Figure 1.** The experimental diffraction pattern of P-IV at 125 GPa (from Ref.2). The tick marks show the peak positions for main (upper) and modulated (lower) reflections. The left inset shows structural model of P-IV in *ac* plane. The basic cell is shown by solid (blue) line. The commensurate approximant with 11-fold supercell along the c-axis is shown by dashed (red) line. A modulation wave 3/11 is shown below. The right inset shows reflections near $2k_F$ indicated by arrow. Indices for reflections are given for the incommensurate cell (4D).

We selected a group of reflections near $2k_F$ for construction of the Brillouin-Jones zone. At first step we constructed planes of reflections (20*lm*) type and the (0012) reflection as shown in Figure 2. On projection down b* one can see a set of planes lying close to the Fermi sphere. This configuration implies that some deformation of the sphere to an ellipsoid would be necessary. Within the free electron model it was suggested to consider a ratio of the Fermi radius to the ½*q* (a distance to a plane from the origin) called as "truncation" factor [7] which is usually 1.05 and may increase up to 1.10 as in the case of (2000) plane for phosphorous.

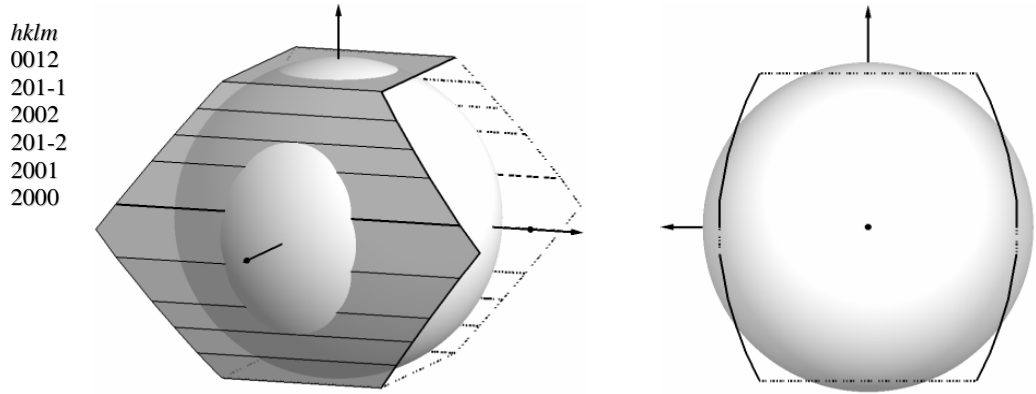

*hklm*
0012
201-1
2002
201-2
2001
2000

**Figure 2.** Configuration of the Brillouin planes (20*lm*) type and (0012) with the inscribed Fermi sphere for the valence electron number z = 5: (left) common view; (right) the view down b*.

Close to the Fermi sphere are planes (1111) and (0210) that are added to the Brillouin – Jones zone shown in Figure 3. This zone is filled to 86 % by electron states which satisfies the Hume-Rothery mechanism of phase stabilization [5].

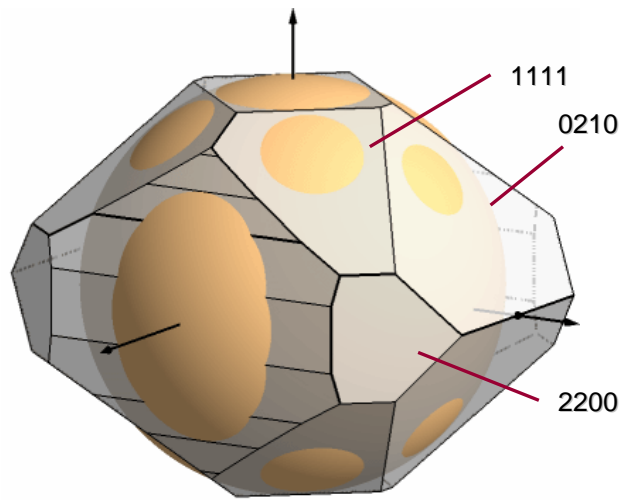

1111
0210
2200

**Figure 3.** The Brillouin – Jones zone for P-IV constructed with planes close to $2k_F$, shown in the inset to Figure 1. Planes (1111), (0210) and (2200) are added to the planes shown in Figure 2.

Now we consider Fermi sphere − Brillouin zone configuration to find factors defining magnitude of the modulation wave vector. It was noted in Ref. [6] that the $2k_F$ vector is matching the (201-1) peak on the diffraction pattern. In Figure 4 we construct the Brillouin planes (001) and (200) in projection down b* and mark nodes (001), (200) and (201) of the basic cell. Constructing the $2k_F$ vector through the point where the (200) plane crosses the Fermi sphere we define the position of the (201-1) reflection. The shift of this reflection down the c* axis from (201) is equal to $\gamma c^*$ and the value of $\gamma$ determined from this construction is 0.268 which is very close to the experimentally found value $\gamma = 0.2673$ [2].

Thus, the plane (201-1) corresponding to the modulation peak (201-1) appears just to cut an empty section where the (200) plane crosses the Fermi sphere. This mechanism is similar to the "nesting" effect and allowed the lowering of electron energy on a Brillouin plane close to the Fermi sphere. Formation of the modulation wave accounts for appearance of other modulation peaks and consequently a group of Brillouin planes near the Fermi sphere as shown in Figures 2 and 3.

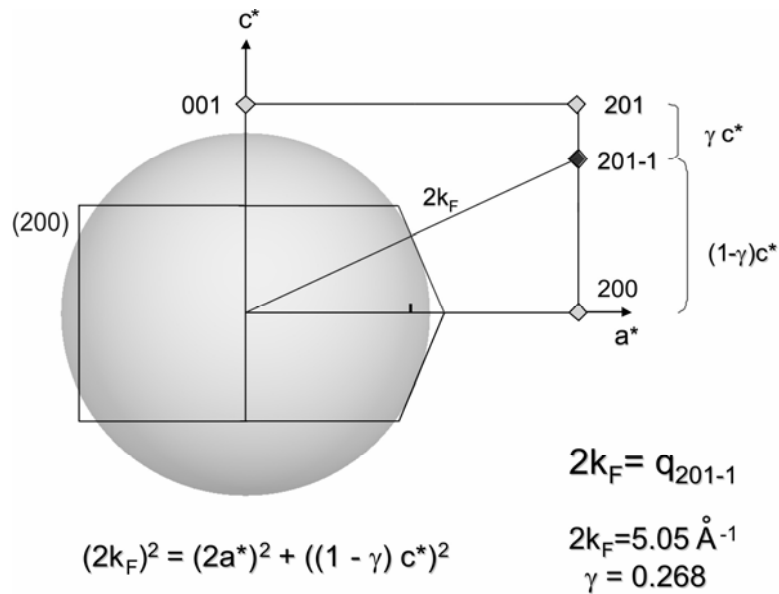

**Figure 4.** Configuration of the rreciprocal lattice for P-IV and the Fermi sphere down b*. Three-integer indices correspond to the basic unit cell. The position of the (201-1) reflection is defined by a reciprocal vector equal to $2k_F$ through the point of crossing of the (200) plane with the Fermi sphere. From this construction one can estimate the value of the modulation wave vector $\gamma$.

It is interesting to compare formation of the modulated structure P-IV to the supercell structure in the CuAu alloy [7]. A common feature is appearance of Brillouin planes close to the Fermi sphere. In the CuAu, an already existing reflection is adjusted to the Fermi sphere by the formation of the commensurately modulated structure with a shift by 1/2 of the period in each 5 cell blocks of a 10-fold supercell. In P-IV structure, the commensurate approximant is nearly as large as in CuAu - an 11-fold supercell. However, the plane lying on the Fermi sphere that defines the modulation is a newly formed one. The modulation wave for the positional displacement of atoms produces a group of Brillouin planes close to the Fermi sphere leading to a more significant effect on the electronic energy.

In conclusion, the stability of the IM structure in P-IV is attributed to the lowering of the electronic band structure energy due to Brillouin zone − Fermi surface interaction. Similar approach can also be applied to the IM structure of the group VI elements.

**Acknowledgments**

Author wish to acknowledge assistance from Dr. Olga Degtyareva and financial support from the Russian Foundation for Basic Research, grant No. 07-02-00901.